\documentclass[twocolumn,aps]{revtex4-1}

\usepackage{graphicx}
\usepackage{amssymb}
\usepackage{amsfonts}
\usepackage{amsmath}
\usepackage{amsbsy}
\usepackage{natbib}
\usepackage{comment}

\usepackage{color}
\usepackage{float}

\usepackage[colorlinks=true, urlcolor=blue, citecolor=blue, linkcolor=blue]{hyperref}

\renewcommand{\vec}[1]{\mbox{\boldmath$\mathrm{#1}$}}
\let\sb=_ \catcode`\_=\active \def_#1{\ensuremath \sb{\rm#1}}

\renewcommand{\vec}[1]{\mbox{\boldmath$\mathrm{#1}$}}
\newcommand{\be}{\begin{equation}}
\newcommand{\ee}{\end{equation}}
\newcommand{\ben}{\begin{eqnarray}}
\newcommand{\een}{\end{eqnarray}}

\begin{document}


\title{The optical tweezer of ferroelectric skyrmions }

\author{X.-G. Wang$^{1}$, L. Chotorlishvili$^2$, V. K. Dugaev$^3$, A. Ernst$^{4,5}$, I. Maznichenko$^{4}$, N. Arnold$^4$, Chenglong Jia$^7$, J. Berakdar$^2$, I. Mertig$^2$, and J. Barna\'s$^7$}
\address{$^1$ School of Physics and Electronics, Central South University, Changsha 410083, China \\
$^2$ Institut f\"ur Physik, Martin-Luther Universit\"at Halle-Wittenberg, D-06120 Halle/Saale, Germany
$^3$ Department of Physics and Medical Engineering, Rzesz\'ow University of Technology, 35-959 Rzesz\'ow, Poland\\
$^4$ Institute for Theoretical Physics, Johannes Kepler University, Altenberger Stra\ss e 69, 4040 Linz, Austria\\
$^5$ Max Planck Institute of Microstructure Physics, Weinberg 2, D-06120 Halle, Germany\\
$^6$ Key Laboratory for Magnetism and Magnetic Materials of the Ministry of Education, Lanzhou University, Lanzhou 730000, China\\
$^7$ Faculty of Physics, Adam Mickiewicz University, 61-614 Pozna\'n, Poland}

\date{\today}

\begin{abstract}

Strong magneto-electric coupling in two-dimensional helical materials leads to a peculiar type of topologically protected solutions -- skyrmions. Coupling between the net ferroelectric polarization and magnetization allows control of the magnetic texture with an external electric field. In this  work we propose the model of optical tweezer -- a particular configuration of an external electric field and Gaussian laser beam that can trap or release the skyrmions in a highly controlled manner. Functionality of such a tweezer is visualized by micromagnetic simulations and model analysis.

\end{abstract}

\maketitle

Optimal dynamical control of a particle motion includes several tasks, such as acceleration, braking, and trapping. In case of nanoparticles, ions, or atoms, the trapping problem becomes more demanding than the others, except trapping of charged particles which is relatively easy with the use of  Pauli trap \cite{Sauter, Diedrich}. In the early 90-ties it was realized that light--atom interaction allows trapping of neutral objects -- cesium and sodium atoms in particular \cite{Davis, Verkerk}. In case of optical trapping of neutral objects, the light does two jobs: (i) it attracts the particles around the nodal points of the optical lattice with the spatial period of the order of  optical wavelength, and (ii) the light additionally cools down the atoms.
The invention of optical tweezers in 1986 by Arthur Ashkin was a triumph for manipulation of microparticles with laser light \cite{Ashkin}. While trapping of various particles is widely discussed in the literature, the problem of trapping of localized excited modes, especially of topological solitons (skyrmions) has not been studied yet.

The concept of skyrmion traces back to the paper of Skyrme \cite{Skyrme}, and to the monumental paper of Belavin and Polyakov \cite{Polyakov}.
It is now well known that skyrmion has a topological character.
In particular, invariance of the topological action of the field theory, $S_{top}\big(\vec{n}\big)=\frac{i\theta}{4\pi}\int dx_{1}dx_{2}\vec{n} \cdot\big(\partial_{1}\vec{n}\times\partial_{2}\vec{n}\big)$,
with respect to the infinitesimal transformation $\vec{n\big(\vec{x}\big)} \rightarrow \vec{n\big(\vec{x}\big)}+\epsilon^{a}\big(\vec{x}\big)R^{a}\vec{n\big(\vec{x}\big)}$, (where $\epsilon^{a}$ is infinitesimal parameter and $R^{a}$ stands for
generators of the O(3) group) defines specific texture of the vector field $\vec{n}\big(x\big)$ \cite{Binz,Dai}.
The set of different textures of $\vec{n}\big(x\big)$, obtained
from each other by means of the continuous deformation, has the same invariant topological action
and the related conserved topological charge $W=\frac{1}{i\theta}S_{top}\big(\vec{n}\big)$. Thus one could argue that the topological soliton (skyrmion) is a robust and
protected object against small perturbations.
Apart from this,  skyrmions possess dual field-particle properties
\cite{Binz,Dai, Iwasaki, HoonHan, Bogdanov, Garst, Gavilano, Bulaevskii, Kong, Mishra, Batista, Hoogdalem, Papanicolaou, Saxena, Rosch}.

Skyrmions are highly mobile objects. There are several precise recipes on how to drive a skyrmion --  either by a spin-polarized electron current or with a magnonic spin current that exerts a magnon pressure on the skyrmion surface.  In the recent work \cite{Xi-guang Wang} an alternative mechanism of skyrmion drag was proposed, which is based on a combination of  uniform temperature profile and non-uniform electric field. Nevertheless, a vital question that arises is whether the particle nature of skyrmions facilitates their trapping.
In what follows, we explore trapping of a skyrmion in the laser field $\vec{e}_p E_{ls}(x,y,z,t)$ (with $\vec{e}_p$ being the unit polarization vector\cite{NikitaArnold}) and the external electric field $ \vec{E}_0 = (0, 0, E_{z0}) $.

Skyrmions emerge in materials (e.g. in chiral single phase multiferroics \cite{Mostovoy}) with a sizeable magnetoelectric (ME) coupling term, $ E_{me} = - \vec{E} \cdot \vec{P} $, where $\vec{P} =  c_{E} [(\vec{m} \cdot \nabla) \vec{m} - \vec{m} (\nabla \cdot \vec{m}) ]$ is the net ferroelectric polarization, with $ \vec{m} $ denoting the unit vector along the magnetization and $ c_E $
 the magneto-electric coupling constant. In chiral multiferroics, coupling of the external electric field with the ferroelectric polarization mimics the Dzyaloshinskii-Moriya term and leads to the noncolinear topological magnetic order. The mechanism of trapping of a skyrmion relies on the interaction between the electric component of the laser field and the ferroelectric polarization of the skyrmion texture.

The laser manipulated skyrmion dynamics is governed by the Landau-Lifshitz-Gilbert (LLG) equation, supplemented by the ME term
\begin{equation}
\displaystyle \frac{\partial \vec{M}}{\partial t} = - \gamma \vec{M} \times \bigg(\vec{H}_{\mathrm{\rm eff}} - \frac{1}{\mu_0 M_s } \frac{\delta E_{me}}{\delta \vec{m}}\bigg)+ \frac{\alpha}{M_{s}} \vec{M} \times \frac{\partial \vec{M}}{\partial t}.
\label{LLG}
\end{equation}
Here, $\vec{M} =  M_{s}\vec{m} $, where $M_{s}$ is the saturation magnetization,  $ \gamma $ is the gyromagnetic ratio, and $ \alpha $ is the phenomenological Gilbert damping constant. The effective field $ \vec{H}_{\mathrm{eff}} $ consists of the exchange field and of the applied external magnetic field, $ \vec{H}_{\mathrm{\rm eff}} = \frac{2 A_{ex}}{\mu_0 M_s} \nabla^2 \vec{m} + H_z \vec{z} $, where $ A_{ex} $ is the exchange stiffness, $ H_z $ is the external magnetic field applied along the \textit{z}-direction.

The z component, $E_{z0}$, of the external electric field stabilizes the skyrmion structure. Due to the Gaussian profile of the electric field component $ E_{ls}(x,y,z,t) $ in the laser beam, the total $z$ component of the electric field, $E_z = E_{z0} + E_{ls}(x,y,z,t)$, is not homogeneous in the $(x,y)$  plane.
Depending on the sign of the oscillating laser field $E_{ls}(x,y,z,t)$, the total field $E_z$ can be either negative or positive.
We note that for an ultrashort laser pulse, the pulse compressor allows control of the spectral phase $\phi(\omega),~E_{ls}(x,y,z,\omega)=\sqrt{|E_{ls}|^{2}}\exp(-i\phi(\omega))$,
where $\phi(\omega)=-\frac{\omega}{c}n(\omega)d$, $n(\omega)$ is the index of refraction and $d$ is the film thickness \cite{Hillman}.
In what follows, we consider both negative $E_z<0$ and positive $E_z>0$ values of the field.
We note that modern laser technologies allow generation of ultrashort single $E_{ls}(x,y,z,t)=E_{ls}(x,y,z)f_{scp}(t)$ and half cycle $E_{ls}(x,y,z,t)=E_{ls}(x,y,z)f_{hcp}(t)$ pulses \cite{Moskalenko}. The temporal profiles of laser pulses are defined as follows: $f_{scp}(t)=t/\tau_{d}\exp(-t^{2}/\tau_{d}^{2})$,
$f_{hcp}(t)=t/\tau_{0}\big[\exp(-t^{2}/2\tau_{0}^{2})-\frac{1}{b^{^{2}}}\exp(-t^{2}/b\tau_{0})\big],t>0$. The ultrashort single pulse has both positive and negative $E_{ls}(x,y,z,t)$, while the negative field part of $f_{hcp}(t)$ is too small. Therefore for half cycle pulse $E_{ls}(x,y,z,t)$ can be viewed as positively defined.

\begin{figure}
	\includegraphics[width=0.45\textwidth]{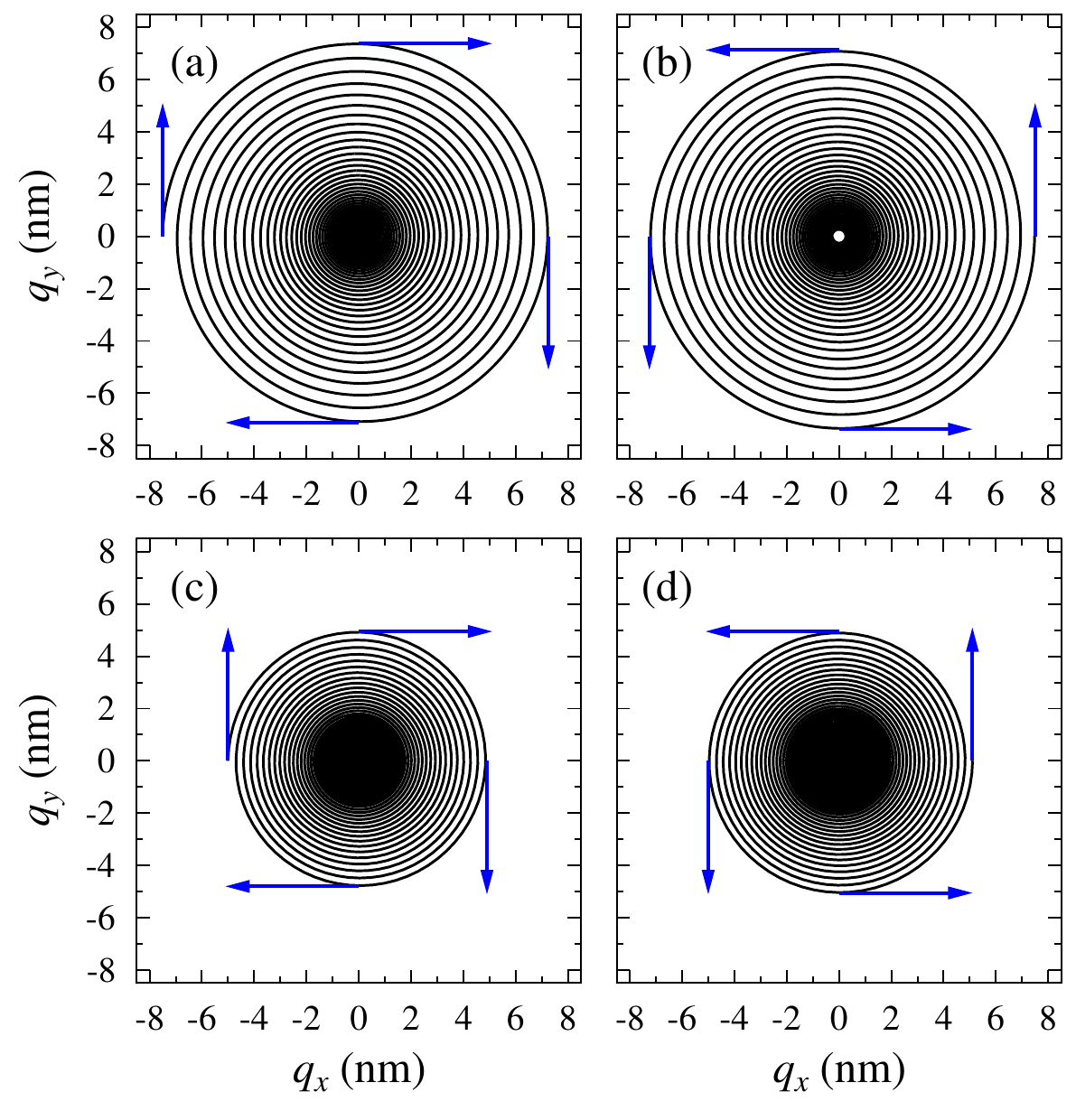}
	\caption{\label{process} (a) The spiral trajectories of the skyrmion winding on clockwise to the laser center. The laser electric field $ E_{0} = 1.2 $ MV/cm. Initially, the skyrmion center $ (q_x,q_y) $ is embedded in the point $ (-7.5,0) $ nm.
(b) The spiral trajectories of the skyrmion winding off anticlockwise from the laser center. The laser electric field $ E_{0} =- 1.2 $ MV/cm.
 Initially, the skyrmion center $ (q_x,q_y) $ is embedded in the point $ (-0.25,0) $ nm.	Numerical solution of the Thiele equation. The spiral trajectory of the skyrmion winding on (c) and off (d) the center of the laser beam. The laser electric field  $ E_{0} = 1.2 $ MV/cm (c) and  $ E_{0} = -1.2 $ MV/cm (d).
 }
\end{figure}
Before presenting the numerical results we explain the trapping mechanism. For the sake of simplicity let us assume that
the electric field is inhomogeneous only in the $x$ direction. The functional derivative of the ME term with respect to the magnetic moment reads:
$-\frac{1}{\mu_0 M_s}\frac{\delta E_{me}(E_i)}{ \delta \vec{m}} = \frac{c_E}{\mu_0 M_s} [\partial_x E_i (m_i \vec{e}_x-m_x\vec{e}_i)
+ \sum_{j(j \ne i)}2E_i(-\partial_j m_j \vec{e}_i + \partial_j m_i \vec{e}_j)].$
Here $ i, j = x, y, z$. We focus on the first term fueled by the nonuniform electric field $\partial_x \vec{E}$, while the second term corresponds to the effective DM interaction with a strength tunable by a constant electric field \cite{Xi-guang Wang}. For tweezing, we suggest using the scanning near-field optical microscopy (SNOM) and advanced nanofabrication procedures. These two methods permit to obtain spots of light $10\sim20$ nm in size; see recent review and references therein \cite{optical fibers}.
Contribution of the nonuniform electric field will be presented in the form of inhomogeneous electric torque (IET):
$-\gamma \vec{m} \times \bigg(-\frac{\delta E_{me}(\partial_x E_i)}{\mu_0 M_s \delta \vec{m}}\bigg) = -\frac{\gamma c_E \partial_xE_i}{\mu_0 M_s} \vec{m} \times (\vec{m} \times \vec{p_E}).$
The vector $ \vec{p_E} = \vec{x} \times \vec{e}_i$ is set by $ \vec{e}_i $ which points into the direction of electric field.
Obviously  the expression of IET is identical to the standard spin transfer torque $ -c_j \vec{m} \times (\vec{m} \times \vec{p}) $,  because $\vec{p_E} $ in IET mimics the spin polarization direction $\vec{p}$. However while $c_j$ depends on the electric current density, the amplitude of the IET depends on the gradient of the electric field $\partial_x E_i$ and on the ME coupling strength $c_{E}$.
In the case of Gaussian laser beam (for more details we refer to the supplementary material), the coefficient in the expression for IET, $ c = \frac{\gamma c_E \partial_r E_z}{\mu_0 M_s} $,  is determined by the gradient of electric field, while $ \vec{p}_E = \vec{e}_r \times \vec{z} $, where $ \vec{e}_r = (\vec{e}_x x+\vec{e}_y y )/\sqrt{x^2 + y^2}  $ is the unit vector. The underlaying mechanism of the skyrmion tweezer is as follows: depending on the direction of the laser field, the IET torque is either centripetal (drives the skyrmion to the center of the beam) or  counter-centripetal (drives the skyrmion out of the beam center).

\begin{figure}
	\includegraphics[width=0.45\textwidth]{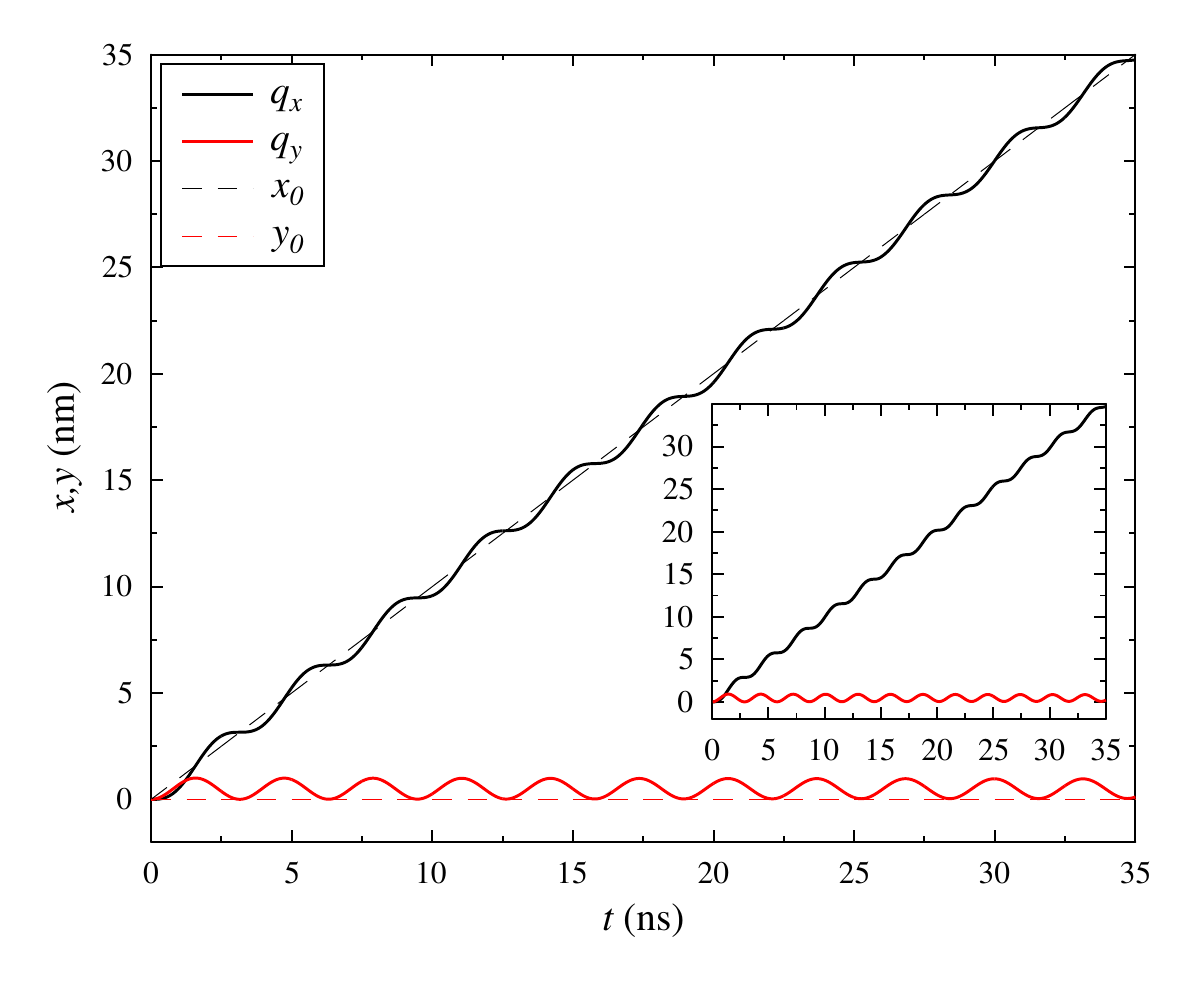}
	\caption{\label{+ez1ms} The center of the laser beam $ (x_0,y_0) $ is steered with the velocity $ v_x $ = 1 m/s and $ E_{0} = 1.2 $ MV/cm. The skyrmion center $ (q_x, q_y) $ trapped by the laser beam follows the motion of the center of the laser beam (black and red colors). The inset plot
shows the numerical solution of the Thiele equations  (see supplementary material). Center of the Gaussian laser beam   $ E_{\mathrm{ls}} $ is steered with the velocity $ v_x $ = 1 m/s and $ E_{0} = 1.2 $ MV/cm.  The skyrmion center $ (q_x, q_y) $ follows the motion of the laser beam center (blue and magenta colors).}
\end{figure}

The numerical simulations based on Eq. (\ref{LLG}) have been done for the saturation magnetization $ M_s = 1.4 \times 10^5$ A/m, the exchange constant $ A_{ex} = 3 \times 10^{-12} $ J/m, the ME coupling strength $ c_E = 0.9 $ pC/m, and the Gilbert damping constant $ \alpha = 0.001$. The N$ \mathrm{\acute{e}} $el-type skyrmion is stabilized by the electric and magnetic fields, $ E_{z0} = 1.7 $ MV/cm  and $ H_{z0} = 4 \times 10^5 $ A/m. In Fig. \ref{process}
we illustrate attraction and repulsion mechanisms of the skyrmion tweezer.
In the first case, Fig. \ref{process} (a), the skyrmion is initially  embedded at the point $ (x,y) = (-7.5,0) $ nm, and  the laser field is positive,  $ E_{ls}(t) > 0 $. Therefore, $ \vec{p}_E = -\vec{y} $,  $ c < 0 $, and the torque  winds the skyrmion on clockwise to the laser beam center $(0,0) $.  In the second case, Fig. \ref{process} (b),  direction of the laser field and IET are reversed, $ E_{ls}(t) < 0 $,  $ \vec{p}_E = \vec{y} $,  $ c > 0 $, and the skyrmion winds out anticlockwise from the laser beam center.
\begin{figure}
	\includegraphics[width=0.45\textwidth]{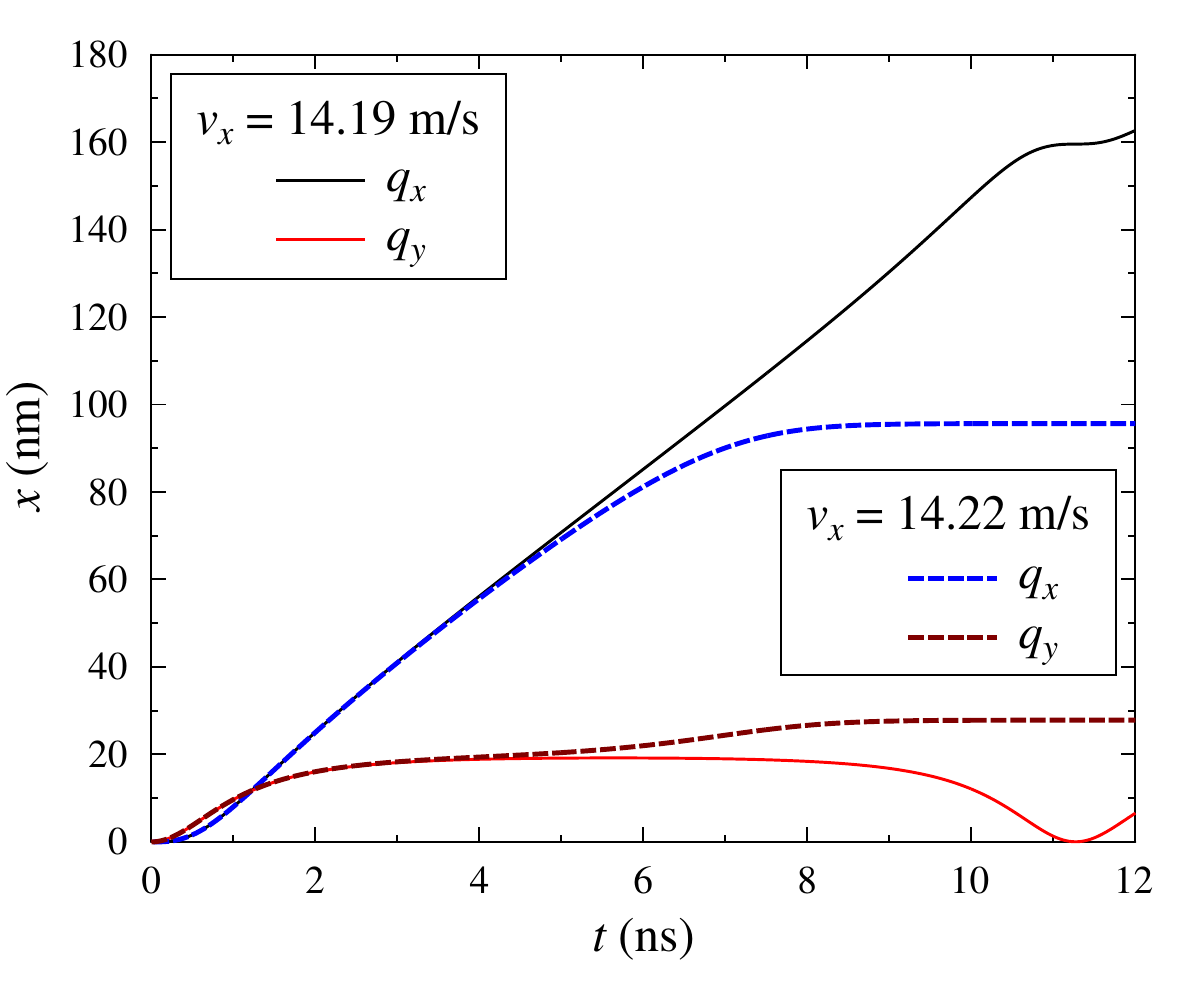}
	\caption{\label{+ezcri} The center of the laser beam is steered with velocity $ v_x $ = 14.19 (14.22) m/s and $ E_{0} = 1.2 $ MV/cm. For $ v_x $ = 14.19 m/s the skyrmion center $ (q_x(t), q_y(t)) $ follows the center of the laser beam. When, $ v_x $ = 14.22 m/s, the skyrmion center is able to follow the laser beam only at the beginning of the evolution, $t <$ 5 ns. }
\end{figure}

\begin{figure}
    \includegraphics[width=0.45\textwidth]{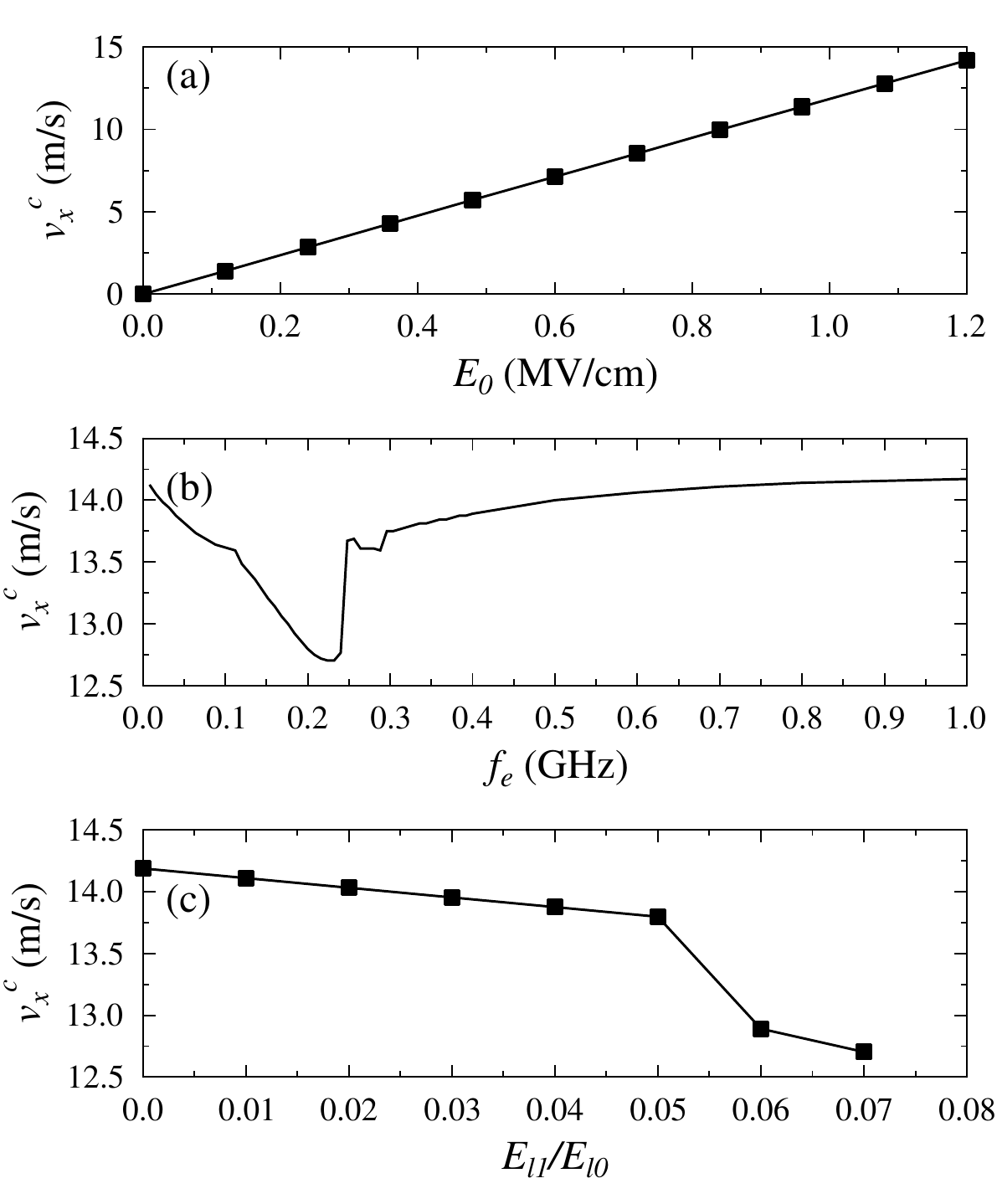}
    \caption{\label{cirtical-fre}
     (a) The critical velocity $ v_x^c $ as a function of the laser electric field $ E_0 $.
     (b) The critical velocity  $ v_x^c $ as a function of the frequency $ f_e $ plotted for electric fields: $ E_{ls} = E_{l0} + E_{l1} \sin(2 \pi f_e t) $, with $ E_{l0} = 1.2 $ MV/cm and $ E_{l1} = 0.07 E_{l0} $. (c) Dependence of the critical velocity on the $ E_{l1}/E_{l0} $ at the frequency $ f_e = 0.22 $ GHz (c).}
\end{figure}

We propose an experimentally feasible strategy for trapping of skyrmions: Focus the laser beam on the center of the skyrmion texture.  Steer the center of the beam  until the electric field is positive $E_{ls}>0$, the skyrmion follows then the center of the beam, see  Fig. \ref{+ez1ms}.  Rotation of skyrmion leads to a weak oscillation of the skyrmion center $ (q_x(t), q_y(t) $. When the beam velocity $v_x$ is  below a critical velocity $v_x^c$, $ v_x < v_x^c = 14.22 $ m/s, increasing of the beam velocity $ v_x $ leads to an increase in the velocity of skyrmion drag. When the beam velocity is above  $  v_x^c $, the skyrmion is not able to follow the center of the laser beam, see Fig. \ref{+ezcri}. The critical velocity $ v_x^c $ increases linearly with $ E_0 $, as is demonstrated in Fig. \ref{cirtical-fre} (a). Thus one can argue that the skyrmion behaves as a massive object. Changing sign of the laser field from positive to negative, $E_{ls}<0$, releases the skyrmion and drives it off the center of the beam (not shown).

We also analyzed the influences of oscillating laser electric field, $ E_{ls} = E_{l0} + E_{l1} \sin(2 \pi f_e t) $. It turns out that the oscillating field drags the skyrmion, and the critical velocity $ v_x^c $  as a function of the frequency $ f_e $ is shown in Fig. \ref{cirtical-fre}(a). The trapping of the skyrmion depends on the frequency of the field. As we see, the critical velocity $ v_x^c $ drops down at  $ f_e = 0.2 $ GHz. Analyzing the spectrum of the skyrmion oscillation frequency (not shown), we find that the frequency $ f_e = 0.22 $ GHz coincides with the natural frequency of the laser-induced pinning potential of the skyrmion, i.e., the resonant oscillation frequency of the rigid skyrmion. The resonant amplification of the skyrmion oscillations leads to release of the skyrmion, and thus reduces $ v_x^c $.
Furthermore, increase of $  E_{l1} $ leads to a decrease of $ v_x^c $, see Fig. \ref{cirtical-fre}(b). The large $ E_{l1} $ activates nonlinear effects and dependence of the critical velocity on the frequency is not linear anymore, see Fig. \ref{cirtical-fre}(b) for $ E_{l1} > 0.05 E_{l0} $.
\begin{figure}
    \includegraphics[width=0.45\textwidth]{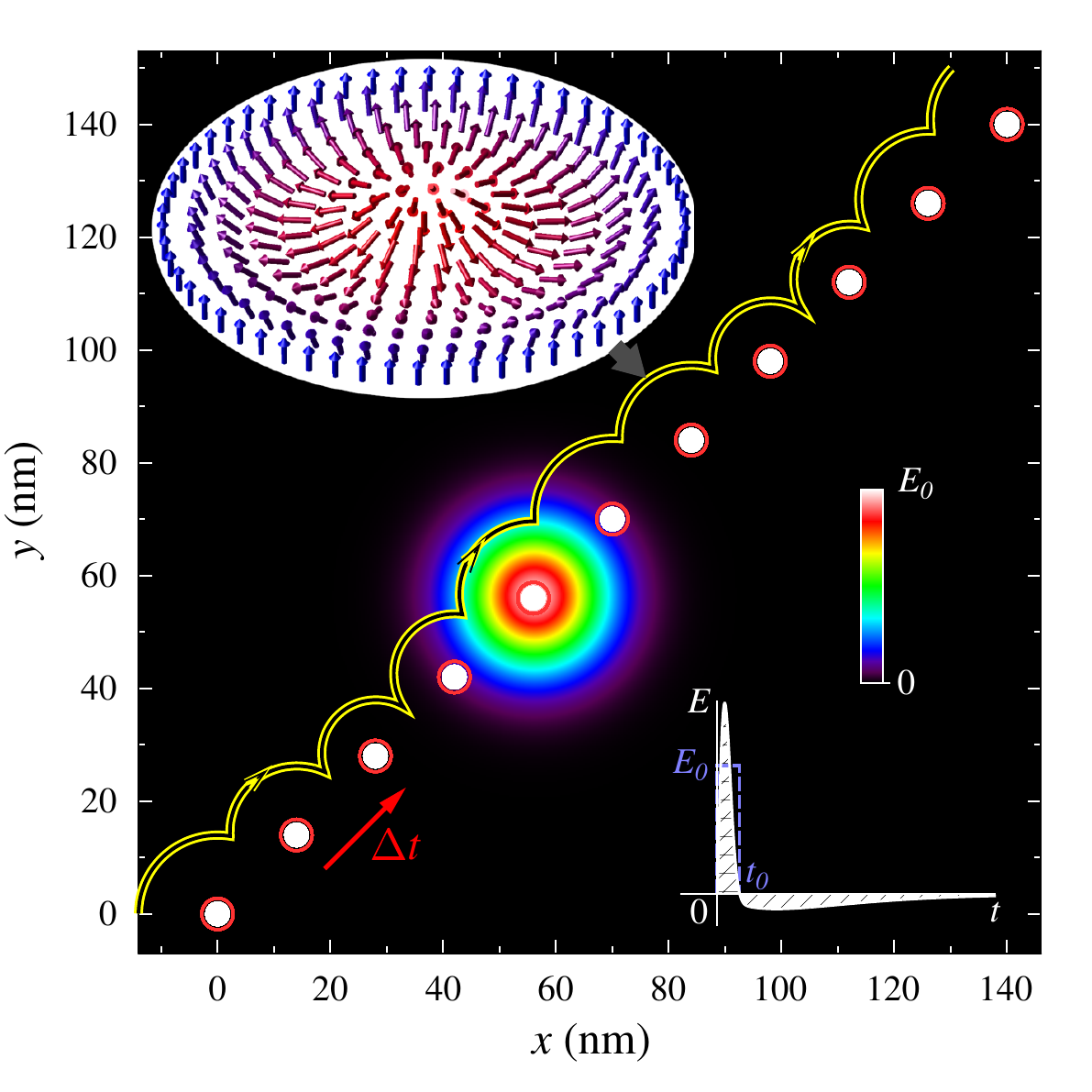}
    \caption{\label{pulse} The skyrmion drag by an oscillating laser pulse.  The skyrmion center $(q_x,~q_y)$ follows the center of the laser beam $(x_0,~y_0)$. The laser center (red dots) is steered in $ 14 \sqrt{2} $ nm in 1.3 ns. For each pulse with whole period 27.3 ns, as demonstrated in inset, $ E_0 = 1.2 MV/cm $ is applied when $ t < t_0 = 1.3 $ ns and it becomes $ -0.06 $ MV/cm for $ t > t_0 $. Right bottom corner: Shape of the
half cycle laser pulse.}
\end{figure}

Akin to the constant Gaussian laser beam, the oscillating laser field also traps the skyrmion. We simulate the laser pulses  70 ps in width and period and steer the center of the laser beam on a distance  $ 14 \sqrt{2} $ nm along $ y = x $ in 1.3 ns. As we see in Fig. \ref{pulse},  the skyrmion is trapped by the laser beam and follows the center of the laser beam (see supplementary material).  The speed of the skyrmion moving along the $y= x $ axis is about 15.5 m/s.
The obtained results can be interpreted in terms of the Thiele equation that describes motion of a rigid skyrmion \cite{zhangcommun10293,Tomasello6784}
Fig. \ref{process} (c), (d), and Fig.(\ref{+ez1ms}) the inset plot.

\begin{figure}
    \includegraphics[width=0.45\textwidth]{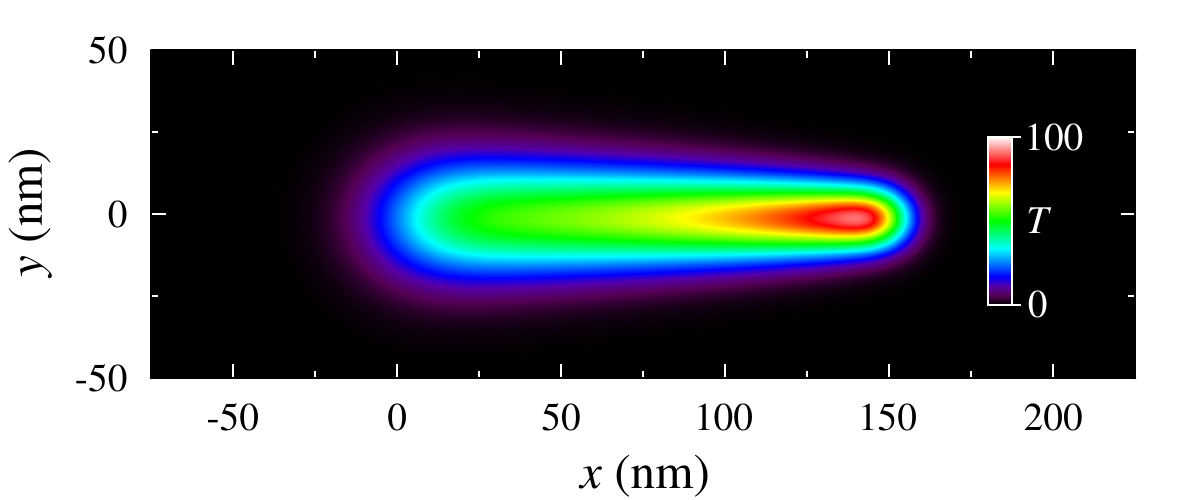}
    \caption{\label{temprofile}  Induced by laser heating temperature profile  $ T(x,y,t) $  at a given time $ t = 11 $ ns. The velocity of the center of laser beam is $ v_x = 14 $ m/s and the amplitude of the electric field $ E_0 $ = 1.2 MV/cm.}
\end{figure}

\begin{figure}
    \includegraphics[width=0.45\textwidth]{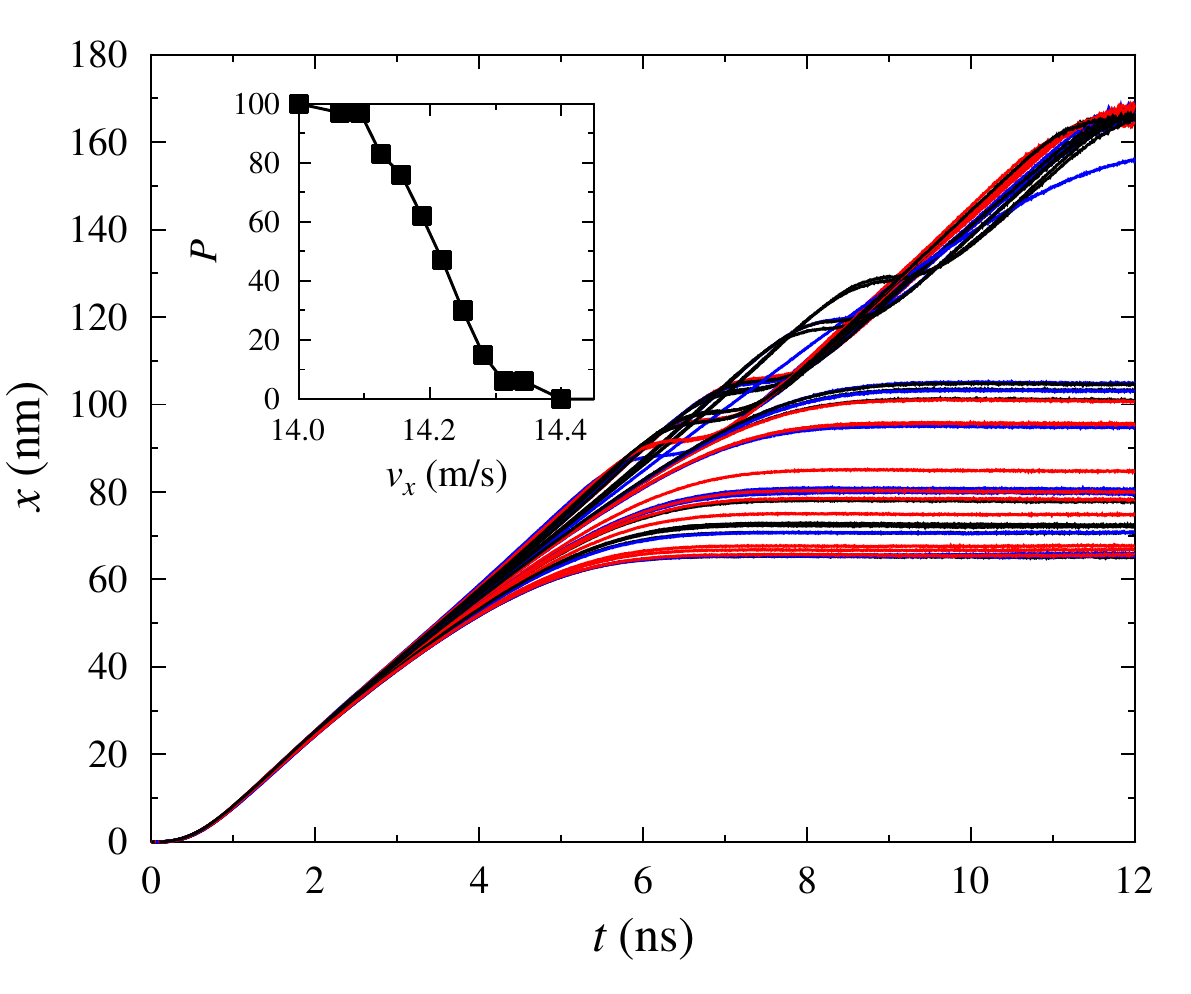}
    \caption{\label{vel14.2-withtem} The effect of the laser heating on the motion of skyrmion. The velocity of the center of laser beam is  $ v_x = 14.22 $ m/s and the amplitude of the electric field $ E_0 = 1.2 $ MV/cm. More than 50 repeated simulations are performed under the same condition. Inset: Dependence of the skyrmion trapping probability  $ P $  on the velocity of the laser beam $ v_x $.}
\end{figure}

A side impact of the laser pulses is the heating effect. The temperature profile $ T(x,y,t) $ induced by the laser heating, through the beam with a moving center is shown in Fig. \ref{temprofile}. The region of the largest temperature (about 100 K) follows the center of the laser beam and the temperature gradually decreases with distance from the center. The laser heating leads to the inhomogeneous time-dependent temperature profile and affects the skyrmion dynamics.  The main effect of laser-induced heating is that the trapping process becomes non-deterministic. We performed a set of calculations with the same initial conditions and collect ensemble statistics see Fig. \ref{vel14.2-withtem}. The trapping probability $ P $ decays for the higher velocity of the center of beam but still is finite. Even at the speed $ v_x = 14.22 $ m/s and $ E_0 = 1.2 $ MV/cm, $ P=44\%$ as is shown in inset in Fig. \ref{vel14.2-withtem}. An interesting fact is that below the threshold velocity of $ v_x \le 14 $ m/s probability $ P = 1 $.
In summary, we have proposed a novel method of optical control of skyrmions in magnetoelectric materials. Owing to the magnetoelectric coupling, electric field of a laser beam couples to the magnetic moments of the skyrmion. When the laser beam is prepared in an appropriate way, one can trap, shift, and then release the skyrmion. Such an optical tweezer may be very useful in optical control and manipulation of the skyrmion position. Numerical results have been obtained from micromagnetic simulations based on Landau-Lifshitz-Gilbert equation with a contribution from magnetoelectric coupling and additionally from solution of the Thiele equation describing motion of rigid skyrmions. A very good agreement of the results obtained by these two methods has been achieved.

\textbf{Acknowledgment:} We are indebted to Albert Fert and Vladimir Chukharev for numerous discussions and suggestions. This work was supported by the National Science Center in Poland as a research project No.~DEC-2017/27/B/ST3/02881, by the DFG through the SFB 762 and SFB-TRR 227, by the National Natural Science Foundation of China No. 11704415,024410-7 and  the  Natural  Science  Foundation  of  Hunan  Province  of China No. 2018JJ3629. A. E. acknowledges financial support from DFG through priority program SPP1666 (Topological Insulators), SFB-TRR227,  and OeAD Grants No. HR 07/2018 and No. PL 03/2018.

\newpage

\section{Supplementary information}

\subsection{Laser pulses}

\begin{figure}
    \includegraphics[width=0.45\textwidth]{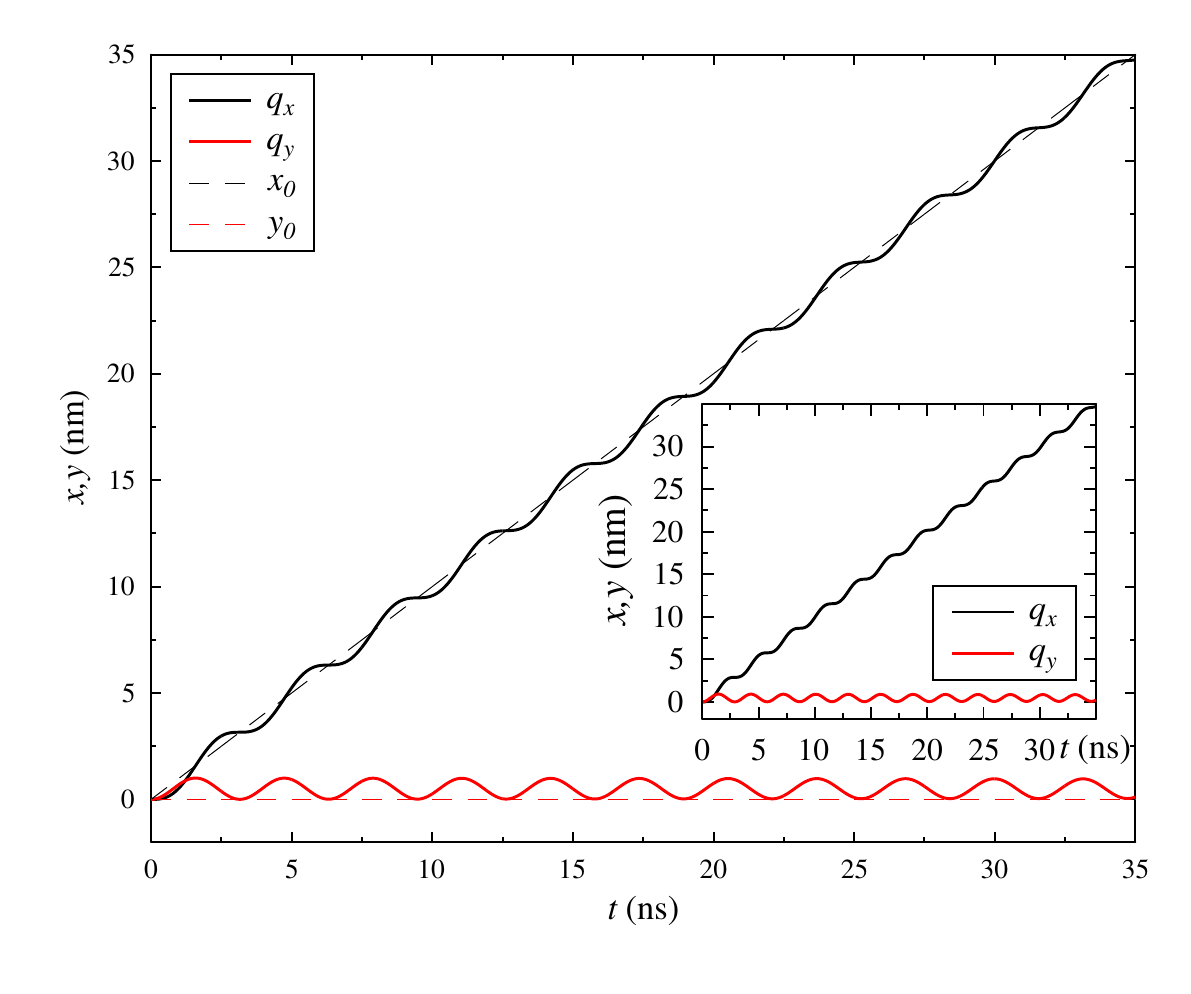}
    \caption{\label{supportingpulse} Time dependence of the skyrmion center $ (q_x,q_y) $ and laser center $(x_0,~y_0)$.}
\end{figure}

Let us assume that the field issued by the laser is a Gaussian beam $ E_{ls}(x,y,z=0,t) =  E_0 f(t) \exp\big[-\frac{(x-x_0)^2+(y-y_0)^2}{\sigma_{0}^2}\big]$ and the distance between laser and film surface is $z_{0}$. After a little algebra one finds the expression for the field at the surface of the skyrmion:
\begin{eqnarray}
&&E_{ls}(x,y,z_{0},t) =  E_0 f(t)\frac{\sigma_{0}}{\sigma}\exp\bigg[-\frac{(x-x_0)^2+(y-y_0)^2}{\sigma^2}\bigg]\nonumber\\
&&\times\exp\bigg[ik\bigg(z_{0}+\frac{(x-x_0)^2+(y-y_0)^2}{2R}\bigg)+i\varphi\bigg].
\end{eqnarray}
Here $f(t)\equiv f(t)_{hcp},~f(t)_{scp}$ is either half or single cycle pulse, $\sigma^{2}(z_{0})=\sigma_{0}^{2}\bigg(1+\bigg(\frac{2z_{0}}{k\sigma_{0}^{2}}\bigg)^{2}\bigg)$ is the width of the beam at the skyrmion surface, $R(z_{0})=z_{0}\bigg[1+\left(\frac{k\sigma_{0}^{2}}{2z_{0}}\right)^{2}\bigg]$, $\tan\varphi=\frac{k\sigma^{2}_{0}}{2 z_{0}}$ and the total electric field acting on the skyrmion has the form $E_{z}=E_{z0}+E_{ls}$. The skyrmion captured by the half cycle laser pulse follows the motion of the beam center see Fig.\ref{supportingpulse}.

Non-paraxial focusing of radially polarized light creates a dominant $E_z$. This effect has a clear interpretation within the framework of geometrical optics. Through the non-paraxial focusing procedure, $E_z$ components of different rays add on the axis, while $E_x,~E_y$ components cancel \cite{Nikita,Youngworth}. We note that narrow laser beam spots can be archived through the shielding of the laser beam by the optical fibers covered by metal. The scanning near-field optical microscopy (SNOM) techniques and advanced nanofabrication procedures allow getting light spots as small as $10\sim20$ nm see recent review and references therein \cite{optical fibers}. Alternatively, in the experiment, one can use a plasmonic tip. In this case, the field as well is nonuniform. However, it is much more complicated, has not Gaussian form, and therefore is less relevant for numerical calculations.

\subsection{Linear and nonlinear tweezing terms}

The tweezing mechanism is based on the inhomogeneous electric torque (IET) \cite{Chotorlishvili}. Here we show that the expression of the IET contains linear and nonlinear in the laser field terms
\begin{equation}\label{IET1}
-\frac{\gamma c_E \partial_xE_i^{laser}}{\mu_0 M_s} \vec{m} \times (\vec{m} \times \vec{p_E}).
\end{equation}
The vector $ \vec{p_E} = \vec{x} \times \vec{e}_i$ is set by $ \vec{e}_i $ which points into the direction of electric field, and the nonlinear magnetic texture in Eq.(\ref{IET1}) is defined by the external electric field $E_{ext}$.
We rewrite Eq.(\ref{IET1}) in the form:
\begin{eqnarray}\label{IET2}
-\frac{\gamma c_E \partial_xE_i^{laser}}{\mu_0 M_s}\vec{i}_{i} \epsilon_{ijk}\epsilon_{klq}m^{j}m^{l}p_{E}^{q}.
\end{eqnarray}
Here $\vec{i}_{i}$ is the unit vector along the axis $i$.
The magnetic texture of the skyrmion is formed by the constant external field and is perturbed by the laser field. This allows us to present the magnetic components in the form:
\begin{eqnarray}\label{IET3}
m^{j}\approx \langle m^{j}\rangle_{ext}+\bigg\langle\frac{\partial m^{j}}{\partial E_{i}^{laser}}\bigg\rangle E_{i}^{laser}.
\end{eqnarray}
After inserting Eq.(\ref{IET3}) into Eq.(\ref{IET2}), similar to A. Ashkin \cite{Tweezer} we obtain not only linear but quadratic terms $\big\langle\frac{\partial m^{j}}{\partial E_{i}^{laser}}\big\rangle\partial_x\big(E_i^{laser}\big)^{2}$.
The nonlinear terms allow the laser field to get stuck with the perturbed magnetic texture and tweeze the skyrmion.

\subsection{The Thiele equation}

Taking into account  Eq. (1), the Thiele equation in our particular case read:
\begin{equation}
\begin{small}
\begin{aligned}
 -\alpha D \partial_t q_x - \partial_t q_y + B e_x &= 0, \\
 \partial_t q_x - \alpha D \partial_t q_y + B e_y &= 0.
\label{Thiele}
\end{aligned}
\end{small}
\end{equation}
Here, $ D \approx 1 $ represents the dissipative force, $ e_x = (q_x - x_0)/\sqrt{(q_x - x_0)^2 + (q_y - x_0)^2} $ is the $ x $ component of $ \vec{e_r} $ located at the skyrmion center $ (q_x, q_y) $, and $ e_y = (q_y - y_0)/\sqrt{(q_x - x_0)^2 + (q_y - x_0)^2} $ is the $ y $ component of $ \vec{e_r} $. The driving force is $ B = - \frac{\gamma c_E \partial_r E_{\mathrm{ls}}}{\mu_0 M_s} L_{sc} I $, where $ L_{sc} $ and $I$ are the scaling length and scaling factor, $ I \approx -0.01 $, we find from micromagnetic simulations. Assuming a constant $ B e_{x,y} $, we deduce steady velocities $ v_x = \frac{\alpha D B e_x - B e_y }{1+\alpha ^2 D} $ and $ v_y = \frac{B e_x + \alpha D B e_y }{1+\alpha ^2 D} $.
Numerical solutions of the Thiele equation Eq. (3) recover the results of the micromagnetic simulations, see in the main text Fig. (1) (c), (d), and Fig.(2).

The temperature profile  $ T(x,y,t) $ is the solution of the heat equation:
\begin{equation}
\displaystyle \frac{\partial T(x,y,t)}{\partial t} = \frac{k_{ph}}{\rho C} \nabla^2 T(x,y,t) + I(x,y,t).
\label{heat equation}
\end{equation}
Here, $ k_{ph} = 0.02 $ W/(m K) is thermal conductivity, $ \rho = 5170 $ kg/m$ ^{3} $ is the mass density, and $ C = 570 $ J/(kg K) is the heat capacity. The source term is $ I(x,y,t) = \frac{l_x\delta_T c \epsilon_0}{2\rho C} E_{ls}^2 $ , where $ c $ is the light speed, $ \epsilon_0 $ is the permittivity of vacuum, $ \delta_T = 1.5 \times 10^6 $ m$ ^{-1} $ is the laser penetration depth, and $ l_x = 1\% $ is the absorption efficiency of laser energy. The temperature effect can be included in the LLG equation Eq. (1) through the random magnetic field $ \vec{h}_{th} $, and its correlation function $ \langle h_{th,i}(t,\vec{r}) h_{th,j} \rangle = \frac{2 k_B T \alpha}{\gamma M_s V} \delta_{ij} \delta(\vec{r}-\vec{r}') \delta(t-t') $. Here, $ k_B $ is the Boltzmann constant, and $ V $ is the volume of the sample.

\end{document}